# Musical Information Extraction from the Singing Voice

*Preeti Rao*

Department of Electrical Engineering, Indian Institute of Technology Bombay
prao@ee.iitb.ac.in

## Abstract

Music information retrieval is currently an active research area that addresses the extraction of musically important information from audio signals, and the applications of such information. The extracted information can be used for search and retrieval of music in recommendation systems, or to aid musicological studies or even in music learning. Sophisticated signal processing techniques are applied to convert low-level acoustic signal properties to musical attributes which are further embedded in a rule-based or statistical classification framework to link with high-level descriptions such as melody, genre, mood and artist type. Vocal music comprises a large and interesting category of music where the lead instrument is the singing voice. The singing voice is more versatile than many musical instruments and therefore poses interesting challenges to information retrieval systems. In this paper, we provide a brief overview of research in vocal music processing followed by a description of related work at IIT Bombay leading to the development of an interface for melody detection of singing voice in polyphony.

## 1. Introduction

Search and retrieval of textual information has attained a fairly mature state as seen by the very widespread use of popular Internet search engines. Multimedia data, of which a vast and growing amount is added to publicly accessible sites each day, however continue to be relatively opaque to search engines. It is not possible to search for music or images unless the multimedia content is tagged with elaborate metadata, typically provided by a manual annotation effort. It is of great practical importance therefore to find automatic methods to extract the relevant metadata by the analyses of the raw audio or video data. In the present paper, we consider the automatic content analysis of audio, music particularly, in order to extract high-level attributes that can be used by search engines. Music information retrieval (MIR), as the field of research is known, actually has a wider scope with the same content analysis methods being useful in musicological studies and in music learning aids.

An early goal of MIR research was to allow users to search for music content by using acoustic queries rather than typed-in text e.g. query-by-singing and query-by-example. The past decade has seen a burst of interest in research in this area fueled by increasingly easy access to large collections of music in digital form [1]. Music recommendation systems can be envisaged around automatically notated genre, artist, style, instrumentation and other musically interesting descriptions of an audio track. A typical approach is to apply signal processing techniques to the extraction of low-level signal features from which mid-level musical descriptors can be derived. Finally a machine learning approach is adopted to link the mid-level descriptions with high-level categories such as genre and artist, or even semantic labels such as mood of the music.

In the present work, we focus on the extraction of melody, an important descriptor of the music, directly applicable in MIR systems such as query-by-singing/humming [2], and key (or even raga) based identification of music. We consider vocal music in particular due to its dominance in popular music as well as the challenges posed by the singing voice to MIR methods. Content analysis of vocal music can potentially provide information about the tune or melody of the song, the lyrics, style and singer. The difficulty of the task is compounded by the presence of accompanying instruments in the polyphonic audio.

An understanding of the signal characteristics of vocal music plays an important role in the design and performance of information extraction systems. In the next section, we describe peculiar signal characteristics of vocal music. In Section III we provide descriptions of recent research related to vocal melody extraction carried out at IIT Bombay.

## 2. Vocal Music Characteristics

As mentioned before the lead instrument in vocal music is the singing voice. It may be construed that singing is just modified speech and so speech processing techniques, which are very well established, may be directly applicable for singing as well. Although there are similarities between speech and singing, there are also several distinct differences. It is therefore instructive to compare singing voice characteristics to those of speech.

As far as similarities are concerned, both speech and singing are comprised of voiced (e.g. vowels) and unvoiced (e.g. stops and fricatives) utterances and voiced utterances usually exhibit predominantly harmonic structure in the frequency domain. There are several differences though [3]. The time ratio for voiced/unvoiced/silent phonation in speech is approximately 60%/25%/15% as compared with the nearly continuous 95% for singing. It is also possible that in singing some vowels may be mutated (modified), deliberately or involuntarily, for projection and/or intelligibility. The average loudness and dynamic range of the singing voice is greater than speech. In singing, specifically operatic singing, there is a tendency for the singer to group the third, fourth and fifth formants together, into what is commonly referred to as the singer's formant, which enhances audibility of the singing voice in the presence of instruments (formants are resonances in the spectral amplitude envelope of the vocal tract).

One of the most important differences between speech and singing is the use and control of pitch. Pitch is a psychological percept and can be defined as perceptual attribute that allows the ordering of sounds in a frequency-related scale from low to high, or more exactly as the frequency of a pure sine tone that is matched to the target sound by human listeners [4]. The physical correlate of pitch is the fundamental frequency (F0), which is defined, for periodic and quasi-periodic sounds only, as the inverse of the time period. The pitch range of normal speech is between 80 and 400 Hz while that of singing can be between 80 and 1000 Hz [5]. Singing also differs from speech in its expressivity, which is physically manifested as instability of its pitch contour. In western singing, especially operatic singing, voice pitch instability is marked by the widespread use of vibrato, a periodic, sinusoidal modulation of phonation frequency during sustained notes [6]. Within non-western forms of music, specifically Indian classical music, voice pitch inflections and ornamentation (*gamaka*) are extensively used as they serve important aesthetic and musicological functions.

In addition to focusing on singing characteristics, techniques for extracting voice-related information from vocal music need to also be robust to signal corruptions caused by the polyphonic (multi-source) nature of music i.e. presence of accompanying musical instruments and/or noise. In the next section we describe our work on melody extraction from vocal music, in which we have made systematic advances into a standard framework, targeted towards exploiting voice characteristics and increasing robustness to interference from accompanying instruments.

## 3. Melody Extraction

The melody of a song is the pitch sequence that a listener might reproduce if asked to hum a segment of polyphonic music [7]. As far as vocal music is concerned, the melody is represented by the pitch contour of the singing voice [8]. A melody extractor therefore detects whether the singing voice is present and estimates the pitch at regular intervals (usually every 10 ms) throughout the duration of the singing voice segments in the polyphonic audio track. The pitch detector has to contend with the non-stationarity of the voice, the presence of unvoiced sounds in singing as well as the interference from the accompanying instruments. All these factors influence the accuracy of the pitch detection adversely and need to be dealt with in order to achieve any useful melody extraction.

The majority of melody extraction algorithms by-and-large adhere to a standard framework, as depicted in Fig.1 [7]. Potential pitch (also referred to as F0) candidates are detected by a pitch detection algorithm (PDA), which may be time-domain based or frequency-domain based. Frequency domain algorithms are based on detecting the regularly spaced harmonics expected in the spectrum of a periodic signal. The assumed smoothness of the pitch contour is used to form a continuous trajectory of the most likely pitch values. To elaborate further, a short-time spectral representation is computed from the input polyphonic audio signal. This is then input to a multi-F0 extraction block whose goal is to detect candidate F0s and associated salience (reliability index) values. The melody identification stage attempts to identify a trajectory through the F0 candidate-time space that most likely represents the melody of the song. The voicing detection block identifies whether the melody is active or silent at each time instant.

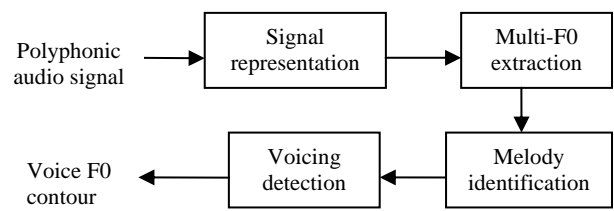

Fig. 1. Block diagram of typical melody extraction system

Our melody extraction system too adheres to the above framework but certain design considerations were specifically tailored for operation on north Indian classical vocal music (NICM) [9], [10]. In NICM large and rapid voice-pitch modulations are frequently present and the accompaniment provided to the vocalist is very characteristic, with a rhythmic percussion that is tonal and relatively strong and a continuously present, relatively weak drone. In addition to performing well on NICM data we found that the design considerations we made also resulted in good performance (on par with state-of-the-art) for western music datasets when evaluated at the Audio Melody Extraction Task at MIREX 2008[1] [11]. We next describe the voice pitch extraction (first three blocks in Fig. 1) and voicing detection mechanism (fourth block in Fig. 1) of our melody extraction system.

### 3.1. Voice Pitch extraction

Here we provide brief descriptions and design considerations made in the signal representation, multi-F0 extraction and melody identification blocks of our system.

#### 3.1.1. Signal representation

It is well known that the frequency-domain analysis of a musical note results in a set of harmonics at integer multiples of an F0. Here we extract a sparse signal representation in the form of a set of harmonic/sinusoid frequencies and amplitudes. For a stationary periodic sound, sinusoidal components in the magnitude spectrum will have a well defined frequency representation i.e. the transform of the analysis window used to compute the Fourier transform. We use a measure of closeness of a local spectral peak's shape to the ideal sinusoidal peak, as defined by Griffin and Lim [12], as a criterion for sinusoid detection.

---

[1] Detailed results are available at www.music-ir.org/mirex/2008/index.php/Audio_Melody_Extraction_Results

The above method for sinusoid identification has been chosen over computationally simpler methods such as fixed amplitude thresholding [13] and amplitude-envelope based thresholding [14], keeping the polyphonic context of the audio in mind. The detection of harmonics of the melodic F0 is critical to the overall performance of the melody extraction system. The use of amplitude thresholding methods may miss these harmonics in the vicinity of strong, interfering harmonics from pitched accompaniment. On the other hand, the use of the above main-lobe matching method will detect even relatively weak sinusoidal components while still maintaining a high side-lobe rejection.

*3.1.2. Multi-F0 extraction*

For several trial F0 that populate the F0 search range (here 70 to 1120 Hz spanning 4 octaves), the corresponding salience is computed as the normalized "Two-Way Mismatch" (TWM) error [15]. The TWM PDA falls under the category of harmonic matching (monophonic) PDAs that are based on the frequency domain matching of a measured spectrum (sinusoids from previous module) with an ideal harmonic spectrum. However, unlike typical harmonic matching algorithms that maximize the energy at the expected ideal harmonic locations, the TWM PDA minimizes a spectral mismatch error that is a particular combination of an individual partial's frequency deviation from the ideal harmonic location and its relative strength.

In previous studies we had found that the melodic (voice) F0 candidate was detected with significantly higher salience in the presence of strong, spectrally sparse, tonal interferences on using the TWM PDA, as compared to other harmonic matching or correlation-based PDAs [9], [10]. This was attributed to the dependence of TWM error values on the frequency extent of harmonics as opposed to individual harmonic strength (which is the case with most 'harmonic-sieve' based methods [7].) This has the advantage that F0s belonging to the singing voice, whose harmonics are known to have a large frequency range, are expected to have lower TWM errors i.e. better salience. Common strong (loud) pitched accompaniment such as guitar, piano, and pitched percussion are spectrally sparse, and will have relatively lower salience.

*3.1.3. Melody identification*

In order to identify the melodic F0 trajectory we use an optimal path finding technique, which dynamically combines F0 salience values (also called measurement cost) and smoothness constraints (also called smoothness cost) using dynamic programming (DP) [16]. The use of DP in melodic identification is preferred since it finds the *optimal* path in one computationally-efficient, *global* framework i.e. a black box that outputs a single F0 contour given suitably defined local and smoothness costs. The local measurement cost for each pitch candidate is given by the normalized TWM error of the F0 candidates obtained in the multi-F0 extraction stage. The smoothness cost must reflect the characteristics of typical voice pitch transitions and is designed based on the musical considerations [9].

The above application of DP in melody identification will result in a single output F0 contour. We found that this approach may lead to incorrect melody identification in the presence of pitched accompanying instruments with comparable salience to the voice pitch and/or which are also capable of producing stable-pitch notes (such as most keyed instruments e.g. piano, harmonium) Recent enhancements to the melody identification module, in terms of extending DP to simultaneously tracking the evolution of two, harmonically unrelated, pitch contours, have resulted in increased robustness to such accompaniment [17].

**3.2. Sung segment detection**

It is not necessary for the voicing detection block to follow the voice-pitch extraction process. In fact the voicing detection mechanism can be considered as a stand-alone system that identifies segments of audio in which the singing voice is present or absent. Such singing voice detection (SVD) systems will follow a generic framework as shown in Fig. 2, in which first a set of features is extracted from the signal and these are input into a classifier which labels the corresponding signal segment as vocal or non-vocal.

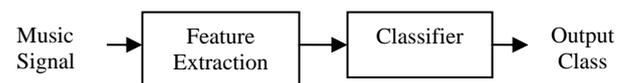

Fig. 2. Block diagram of typical singing voice detection system

Various classifiers, such as Gaussian mixture models (GMM) [18], support vector machines (SVM) [19] and multi-layer perceptrons (MLP) [20]) have been used in previous studies related to SVD but, as Berenzweig, Ellis and Lawrence note, "the methods of statistical pattern recognition can only realize their full power when real-world data are first distilled to their most relevant and essential form" [20]. This emphasizes the importance of the design and selection of features that demonstrate the ability to discriminate between singing voice and accompanying instruments. Commonly used features in previous SVD related studies, such as MFCCs [20], [21], attempt to capture the timbral aspects of musical sounds. However, it is well known that the singing voice occupies a large and diverse timbre-space (due to the continuous variation of vocal tract characteristics with articulation of different phones and also due to variations in vocal tract dimensions across different singers), which may not be completely captured by such features.

In a previous study we have demonstrated the superiority of a voice-pitch based energy feature over previously used standard (timbral) feature sets for the SVD problem, in polyphony, within a standard classification framework (GMM) [22]. However the proposed feature showed a reduced capacity for discrimination between the voice and loud pitched instruments. To increase robustness to such accompaniment we then enhanced the feature by taking into account the temporal variation of the pitch contour, in order to exploit voice-pitch instability (see Section 2), as opposed to keyed- instrument pitch stability. Classification results using this enhanced feature did not show any degradation even in the presence of loud pitched instruments [23].

## 4. Summary and Discussion

As described in the previous sections, a melody extraction system has been developed for the singing voice in polyphonic audio. A number of enhancements to known analysis methods have led to increased robustness of pitch tracking and voice detection functions in the presence of pitched accompanying instruments. The melody extractor has been embedded in an interactive user interface where analysis parameters can be adjusted to obtain the best estimation of the melody as determined by listening to a synthesized version of the detected pitch contour. This facilitates the recovery of the melodic pitch contour even in difficult cases of polyphony with minimal manual effort.

The present work on melody extraction can be extended to facilitate other information retrieval tasks for polyphonic music such as singer identification by using the detected pitch to isolate the singing voice harmonics and hence its spectral envelope. In the same vein, it may be explored for improving lyrics recognition, which remains a difficult problem in state-of-the-art systems.